\documentclass[10pt]{iopart}
\usepackage{graphicx}
\usepackage{dcolumn}
\begin{document}
\title{Numerical Evaluation of the Statistical 
Properties of a Potential Energy Landscape}
\author{E. La Nave\dag, S. Mossa\ddag, F. Sciortino\dag, 
and P. Tartaglia\dag}
\address{\dag Dipartimento di Fisica, INFM and
Center for Statistical Mechanics and Complexity,
Universit\`a di Roma "La Sapienza",
Piazzale Aldo Moro 2, I-00185, Roma, Italy}
\address{\ddag Laboratoire de Physique Theorique des Liquides,
Universite Pierre et Marie Curie , 4 place Jussieu,
Paris 75005, France}
\begin{abstract}
The techniques which allow the numerical evaluation of 
the statistical properties of the potential energy landscape 
for models  of simple liquids are reviewed and critically discussed.
Expressions for the liquid free energy and its
vibrational and configurational components are reported.
Finally, a possible model for the statistical
properties of the landscape, which appears to describe
correctly fragile liquids in the region where equilibrium simulations are
feasible, is discussed.
\end{abstract} 
\pacs{PACS number(s):}
\section{Introduction}
\label{intro}
Understanding the dynamic and thermodynamic 
properties of supercooled liquids is one of the more 
challenging tasks of condensed matter physics
(~\cite{debenedettibook,debenedetti01,mezard01,proceedingH,proceedingT,proceedingS}).
A significant amount of 
experimental~\cite{angell95,torre,gotze-pisa,cummins-pisa},
numerical~\cite{kob-review}, and theoretical 
work ~\cite{mct2,speedy,mct1,mezpar,wolynes} 
is being currently devoted to the understanding of the physics 
of the glass transition and to the associated slowing down 
of the dynamics. In the latest years the study of the topological
structure of the potential landscape (PEL)~\cite{stillinger_pes} 
and the connection between the properties of the PEL and the 
dynamical behavior of glass forming liquids has become 
an active field of research. 
Among the thermodynamic formalisms amenable to
numerical investigation of the PEL properties, a
central role is played by the Inherent Structure (IS) formalism introduced 
by Stillinger and Weber~\cite{stillinger_pes}. Properties of the
potential energy landscape, such as depth, number  
and shape of the basins of the potential energy surface
are calculated and used in the evaluation of the
liquid free energy in supercooled state~\cite{scala00,sastry01,voivod01,otplungo}.
In the IS formalism, the system free energy 
is expressed as a sum of an entropic contribution, 
accounting for the number of the explored basins, 
and a vibrational contribution, expressing the free energy of the 
system when constrained in one  of the basins~\cite{stillinger_pes}.

In this manuscript we review the numerical techniques 
which allow the evaluation of the statistical
properties of the PEL for atomic and molecular systems
 \cite{skt99,scala00,sastry01,starr01,speedyjpc,otplungo}.
The paper is organized as follow:
Sec.~\ref{sec:freeenergy_intro} 
provides a brief introduction to  the inherent structure
formalism, introduced by Stillinger and Weber~\cite{stillinger_pes}.   
Within this formalism, an exact expression for the liquid free energy, 
based on landscape properties can be derived. Sec.~\ref{sec:freeenergy} 
reviews the numerical techniques which allow a precise numerical 
evaluation of the liquid free energy.  
Sec.~\ref{sec:eis} describes the numerical techniques
requested for the evaluation of the inherent structure energies.
Sec.~\ref{sec:fvib} discusses techniques to evaluate 
the vibrational component of the free energy. 
Sec.~\ref{sec:sconf} shows how, from the previous information,  
is possible to evaluate the configurational entropy. 
Sec.~\ref{sec:rem}  discusses a possible
modellization of the statistical properties of the 
landscape, based on the hypothesis of a gaussian distribution of basin's 
depth~\cite{heuer00,sastry01,otplungo},  
and compares the predictions of the model with numerical 
results for a molecular system.
\section{The free energy in the IS formalism}
\label{sec:freeenergy_intro}
In the $IS$ formalism ~\cite{stillinger_pes}, 
the free energy of a supercooled liquid is
expressed  in term of the statistical properties of the  
potential energy landscape (PEL).
The potential energy surface  is partitioned in so-called basins, 
defined as the set of points such that a steepest
descent path ends in the same local minimum. The configuration at the
minimum is called inherent structure ($IS$) and its energy and pressure are 
usually indicated as $e_{IS}$ and $P_{IS}$. 
The  partition function can be expressed as sum of the Boltzmann 
weight over all the basins, i.e., as sum over the basin partition functions. 
As a result, the Helmholtz liquid free energy  $F(T,V)$ 
can be written as~\cite{stillinger_pes}: 
\begin{equation}
F(T,V)= \langle e_{IS}(V,T)\rangle - T S_{conf} (T,V)+ f_{vib} (T,V),
\label{eq:freeenergy}
\end{equation}
where
\begin{itemize}
\item
$\langle e_{IS}\rangle $ is the average energy of the explored local 
minima at temperature $T$ and volume $V$;
\item
$f_{vib}$ is the vibrational free energy, i.e., the  free energy
of the system constrained in one basin, a quantity depending on
the shape of the explored basins;
\item
 $S_{conf}$ is the  configurational entropy, that counts
the number  of explored basins.

\end{itemize}

The  numerical evaluation of $F(T,V)$, $\langle e_{IS}(T,V)\rangle$,
and $f_{vib}(T,V)$ is sufficient for calculating
$S_{conf}$ and, from it, the number of basins  
$\Omega(e_{IS})de_{IS}$ with depth between $e_{IS}$ and $e_{IS}+de_{IS}$.
Indeed, in the thermodynamic limit, $\ln \Omega(e_{IS})$
can be derived  from a  plot of $S_{conf}$ vs $\langle e_{IS}\rangle$ 
(parametric in $T$). This quantity, together with the  
$\langle e_{IS}\rangle$ dependence of $f_{vib}$ , provides    
a precise quantification  of the statistical properties of the landscape.
\section{Numerical evaluation of $F(T,V)$}
\label{sec:freeenergy}
This section describes the numerical techniques used to evaluate
the liquid free energy, based on thermodynamic integration
\cite{skt99,coluzzi,scala00,sastry01,otplungo}.
First,  a path in the $(T,V)$ plane, connecting the ideal gas state
to the desired state point, has to be selected. The selected path must
avoid the liquid-gas first order line. A convenient choice is
a constant temperature path (with $T=T_o$ higher than 
the liquid-gas critical temperature) from infinite volume to the 
desired volume, followed by a constant volume path from $T_o$ down to 
the range of temperature  of interest.

In  the general case of a system of $N$ rigid molecules, the ideal gas
free energy is
\begin{equation}
F_{ig}(T,V,N)=-
N k_B T \left\{ 1+\frac{1}{2}\ln \pi- 
\ln \nu +\ln\left [
\frac{V\sqrt{{\cal A}^3{\cal R}_x {\cal R}_y 
{\cal R}_z}}{N} T^3\right ]\right \},
\label{eq:id-freeenergy}
\end{equation}
where 
${\cal A} \equiv \frac{2\pi m k_B}{h^2}$,
${\cal R}_\mu\equiv\frac{8\pi^2 k_B I_\mu}{h^2}$
-- with  $\mu$ denoting x, y, or z --,
$I_\mu$ is the inertia moment of the molecule
with respect the axis $\mu$, and $\ln \nu$  accounts 
for the molecular symmetry.
In the case of $C_{2v}$ molecules (like water) $\nu=2$,
due to the two possible degenerate
angular orientations of the molecule~\cite{fotocopie}.

\begin{figure}
\centering
\includegraphics[width=.90\textwidth]{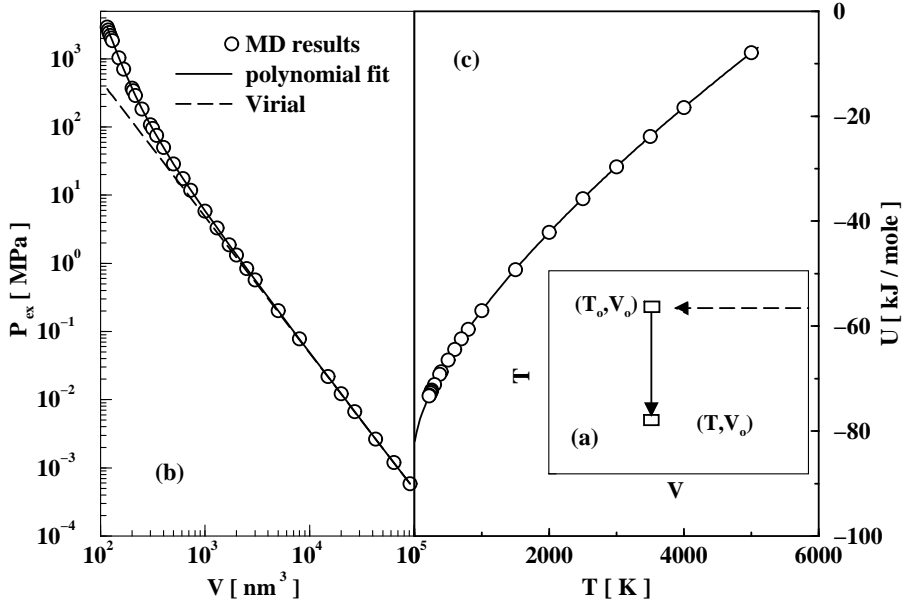}
\caption{(a) Thermodynamic integration paths used to calculate 
the total free energy  at the thermodynamical points of interest,
starting from the ideal ---non interacting-- gas state.
(b) Excess pressure at $T=T_o$  as a function of volume. 
The open circles are the MD results. 
The dashed line is the the first term of the 
virial expansion to the excess pressure; the solid line
is a third order polynomial fit to the entire set of data.
(c) Potential energy (open circles) at the volume $V_o$
in the entire temperature range considered; the solid line 
is the fit of the data. 
The data are from our simulation~\cite{otplungo} of a system $N=343$
molecules modeled by  the  Lewis and Wahnstr\"{o}m  model 
for Orthoterphenyl~\cite{lewis}, whose dynamics~\cite{rinaldi} 
and thermodynamics~\cite{otplungo,press} features have been studied in detail.}
\label{fig1}
\end{figure}

To perform the thermodynamics integration along the isotherm $T_o$, 
one needs to select about 20-30 state points at different volumes 
(Fig.~\ref{fig1}-(a)). Of course, the smallest chosen volume must 
coincides with the final volume $V_o$. The
largest $V$ value $(V_\infty)$ must be chosen in such a way 
that the vast majority of the molecular interactions are binary 
collisions, i.e., that the volume dependence of the pressure 
is well described by the (first order) virial expansion.  At large volumes, 
although dynamics is very fast, care has to be taken to run the 
simulation long enough to sample a large number of binary collisions.

The free energy at $(T_o,V_o)$ can be calculated as
\begin{equation}
F(T_o,V_o)=F_{ig}(T_o,V_o)-\int_\infty^{V_o} dV P_{ex}(T_o,V)+
\frac{U(T_o,V_o)}{T_o},
\label{entropiaco}
\end{equation}
where $U(T_o,V_o)$ is the potential energy and  $P_{ex}(T_o,V)$ is the excess pressure, i.e., the pressure 
in excess to the ideal gas pressure. 
The calculated $P_{ex}(T_o,V)$ curve can be 
fitted according to the polynomial in powers of $V^{-1}$ 
(Fig.~\ref{fig1}-(b)):
\begin{equation}
P_{ex}(T_o,V)=\sum_{k=1}^n a_k(T_o) V^{-(k+1)},
\end{equation}
giving
\begin{equation}
F(T_o,V_o)=F_{ig}(T_o,V_o)
+\sum_{k=1}^n \frac{ a_k(T_o) V_o^{-k}}{k} +\frac{U(T_o,V_o)}{T_o}.
\label{free}
\end{equation}

To perform the thermodynamic integration along a constant $V_o$ path,
it is necessary to evaluate the internal energy $U(T,V_o)$ as a function of 
$T$, from $T_o$ down to the lowest state where equilibration
of the system is feasible (Fig.~\ref{fig1}-(c)).
The resulting free energy $F(T,V_o)$ can be calculated as
\begin{equation}
F(T,V_o)=F(T_o,V_o) + 3 R \; \ln(T/T_o) 
+\int_{T_o}^{T}\frac{dT}{T}\frac{\partial U(T,V_o)}{\partial T}.
\label{eq:freeenergy2}
\end{equation}
The $3 R \ln(T/T_o)$ term accounts for  the ideal gas contribution to the 
free energy. Again, a fit of  $ U(T,V_o)$ vs  $T$ 
is requested to evaluate the integral in the above expression.
One possibility, which has been often found very successful for 
dense systems (small $V_o$)~\cite{coluzzi,skt99}, 
is to fit $ U(T,V_o)$ vs $T$ according to the Tarazona 
law~\cite{tarazona}, i.e., $U(T,V_o)=b_0(V_o)+ b_1(V_o) T^{3/5}$. 
Of course, for the present purposes,
any functional form which correctly represents $U(T,V_o)$ can be
selected.

In summary, performing thermodynamic integration,  
an accurate numerical expression for $F(V,T)$ 
can be obtained.
\section{The average inherent structure energy $\langle e_{IS}\rangle$}
\label{sec:eis}
This section describes how to calculate the
average inherent structure energy $\langle e_{IS} (T,V)\rangle$. 
Recently, it has  been shown that, on cooling at constant volume,  
on entering in the supercooled region,  
the system  starts to explore  
basins  of lower and lower  $e_{IS}$~\cite{sastry00}. 
The $T$ dependence of the average explored basin depth 
follows a $1/T$ law
~\cite{heuer,sastry01,starr01,otplungo}
for fragile liquids. Note that for silica, the prototype of a  
strong liquids, the $1/T$-law is not observed and 
$\langle e_{IS}(T,V)\rangle$ appears  to approach a constant 
value on cooling~\cite{voivod01}. 

In order to evaluate $\langle e_{IS}(T,V)\rangle$
one needs to perform steepest descent potential energy minimizations
for a statistically representative ensemble of equilibrium 
configurations, to locate their corresponding inherent structure,
i.e., local minima. For efficiency reasons, the search for the closest 
local minima is performed using the conjugate gradient 
algorithm~\cite{num-recepies}.  In this algorithm, the system evolves
along a sequence of straight directions until the minimum is reached. 
In each step, the new search direction  recalls the directions 
already explored, improving the algorithm efficiency.
In rigid molecule systems, each step is composed by a
sequence of minimization of the center of mass coordinates,
followed by a minimization of the angular coordinates. 
Rotations around the principal axis of the molecule are often
chosen. The minimization procedure is continued until the 
energy changes less than a preselected precision. 
Since the change in  $\langle e_{IS}(T)\rangle$ in supercooled 
states is often less than  one per cent of its own value, 
a high precision is requested in the minimization procedure. 

\begin{figure}
\centering
\includegraphics[width=.90\textwidth]{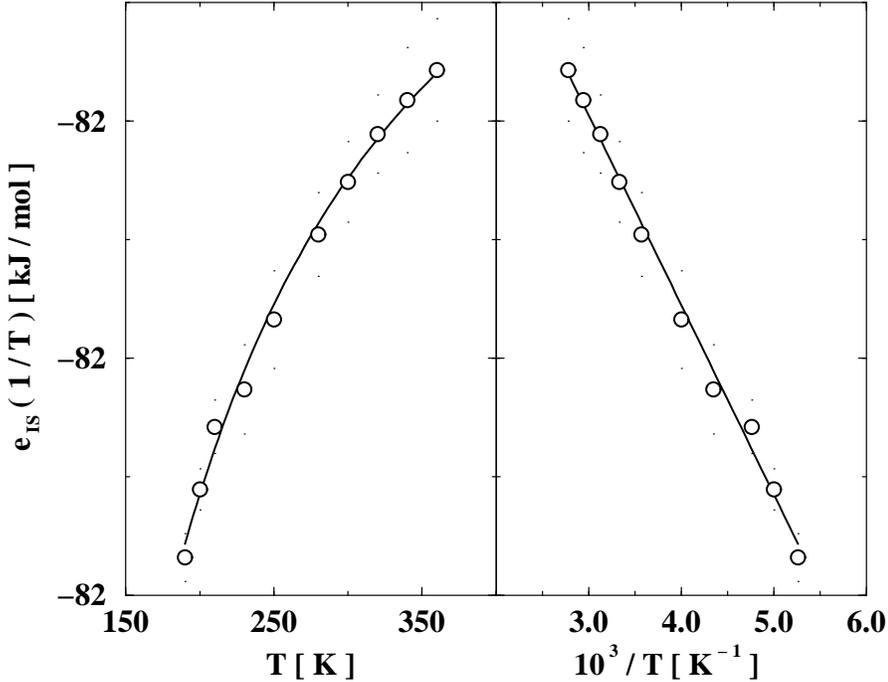}
\caption{ 
$\langle e_{IS}\rangle$ as a function of $T$ (a), and as a function of
$1/T$ (b). Data are from simulations~\protect\cite{otplungo} 
of a system $N=343$ molecules modeled by the Lewis and Wahnstr\"{o}m  
model for Orthoterphenyl~\protect\cite{lewis}.}
\label{fig2}
\end{figure}

In Fig.~\ref{fig2} we show $\langle e_{IS}\rangle$ (a) 
as a function of $T$ and (b) as a function of
$1/T$ for a rigid molecular model.
\section{The vibrational free energy}
\label{sec:fvib}
The vibrational free energy $f_{vib}(e_{IS},T,V)= 
U_{vib}(e_{IS},T,V) -T S_{vib} (e_{IS},T,V)$ 
is the free energy associated to the
exploration of a basin of depth $e_{IS}$ at temperature $T$ 
and volume $V$. $f_{vib}(e_{IS},T,V)$ takes into account both
the kinetic energy of the system and 
the local structure of the basin with energy $e_{IS}$.
~From a formal point of view, it is defined as 
\begin{equation}
f_{vib}(e_{IS},T,V)=-k_B T \ln \left( \frac{\Lambda_x \Lambda_y \Lambda_z}
{\lambda^{3N}} \sum {'}  \frac {\int_{V_{basin}} 
exp ( -\beta [V({\bf r^N}) -e_{IS}]) d {\bf r^N }  } 
{\Omega(e_{IS})de_{IS}} \right),
\label{eq:fvibcorr} 
\end{equation}
where $\sum '$ is the sum on all the basins with energy depth
$e_{IS}$.
The integration of the  Boltzmann factor is performed over all points in
configuration space associated to the selected basin.
Here ${ \Lambda}_\mu \equiv (2\pi I_\mu k_B T)^{1/2}/ h$,
$\lambda\equiv h (2\pi m k_B T)^{-1/2}$ is the de Broglie wavelength, and
$V({\bf r^N})$ is the potential energy. 

The evaluation of the integral requires the exact knowledge of the
shape of the PES in the basin and, in general, it will give rise 
to a complex $T$-dependence of the vibrational energy.
The best that can be done at the present time is to assume that
the $e_{IS}$ dependence in $f_{vib}$ is captured by the
$e_{IS}$ dependence of the density of states of the basin, 
evaluated at the IS configuration~\cite{otplungo,press}. 
In other words, the vibrational free energy is split into
an harmonic contribution (which depends on the
curvature of the potential energy at the minimum) 
and an anharmonic contribution, which is often
assumed basin independent.
In molecular systems the Hessian, the matrix of the 
second derivatives of the potential energy, is calculated numerically  
selecting as molecular coordinates the center of mass and the 
angles associated with the rotations around the three molecular 
principal inertia axis.
Diagonalization is performed with standard numerical routines.

In harmonic approximation, the free energy associated
to a single oscillator at frequency $\omega$ is
$k_BT \ln( \beta \hbar \omega)$. Hence, the basin
free energy can be written as
\begin{equation}
f_{vib}(e_{IS},T,V)=
k_B T \langle \sum_{i=1}^{6N-3} \ln(\beta\hbar\omega_i(e_{IS}))\rangle' 
+ F_{anh}(T,V),
\label{eq:fbasin}
\end{equation}
with
\begin{equation}
U_{vib}(T,V)=(6N-3) \frac {k_B T}{2} + U_{anh}(T,V), 
\end{equation}
and
\begin{equation}
S_{vib}(e_{IS},T,V)=\langle\sum_{i=1}^{6N-3} \left( 1-\ln \left[\frac{\hbar
\omega_i(e_{IS})}{k_BT}\right] \right)\rangle' + S_{anh}(T,V);
\label{eq:svib}
\end{equation}
here the $\omega_i(e_{IS})$ are the frequencies of the 
$6N-3$ independent  harmonic oscillators given by 
the square root of the $6N-3$ non-zero eigenvalues of 
the Hessian matrix evaluated in the IS. 
The  $\langle  \rangle'$  is the average on all the 
basins with the same energy $e_{IS}$.  
Note that the above equations are derived 
assuming:
\begin{equation}
\langle \sum_{i=1}^{6N-3}\ln(\beta\hbar\omega_i(e_{IS}))\rangle'= \ln
 \langle \exp^{\sum_{i=1}^{6N-3}\ln(\beta\hbar\omega_i(e_{IS}))} \rangle'.
\end{equation}
For the molecular systems studied so far,   this unnecessary approximation introduces an error smaller than $1 \%$. The relevant approximation
consists in dropping the $e_{IS}$ dependence in  the anharmonic 
contribution to the  vibrational free
energy $F_{anh}$ (and of course in $U_{anh}(T,V)$ and
$S_{anh}(T,V)$). In other words, the
anharmonicities are assumed to be identical in all basins.
Under such approximation, $U_{anh}(T,V)$  can be calculated from the
simulation data as
\begin{equation}
U_{anh}(T,V)= U (T,V)-\langle e_{IS}(T,V)\rangle-(6N-3) 
\frac{k_B T}{2},
\end{equation}
and it can be well fitted by an expansion in powers of $T$, 
starting from $T^2$, as
\begin{equation}
U_{\rm anh}(T,V)=\sum_{k=2}^{N_c} c_k(V) T^k.
\label{anh_en}
\end{equation}
Correspondingly, $S_{anh}(T,V)$ can be estimated
by thermodynamic integration along the 
isochore between  temperatures $0$ and $T$ as
\begin{equation}
S_{\rm anh}(T,V)=\int_{0}^T \frac{dT'}{T'} \frac{\partial U(T,V)}
{\partial T}=\sum_{k=2}^{N_c} \frac{k c_k(V)}{k-1} T^{k-1}.  
\label{anh_ent}
\end{equation}
By incorporating the anharmonic corrections, which in models of 
simple fragile liquids studied so far are
not particularly significant~\cite{skt99,otplungo}, 
a good estimate of the basin free energy is obtained.  
We note on passing that for the cases of network forming liquids, 
anharmonic corrections are relevant~\cite{starr01,voivod01} 
and must be taken into account.

\begin{figure}
\centering
\includegraphics[angle=0,width=.90\textwidth]{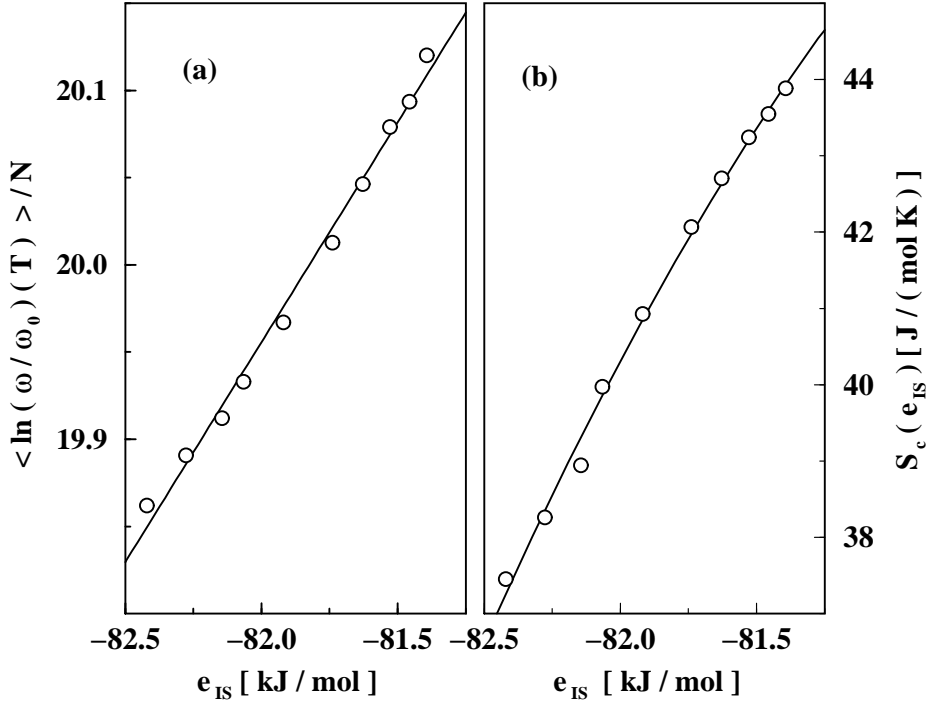}
\caption{${\cal V}$ (a) and  $S_{conf}$ (b) 
as a function of $e_{IS}$. }
\label{fig3}
\end{figure}

It is important to note that all the $e_{IS}$ dependence in the average  
basin free  energy is  carried by the term ${\cal V}(V,e_{IS}) 
\equiv \langle \sum_{i=1}^{6N-3} \ln \omega_i(e_{IS})\rangle$. 
A parameterization of such quantity as a function of $e_{IS}$ 
allows to simply connect the basin free energy to the basin depth. 
In all models studied so far~ \cite{sastry01,otplungo,st01,press}
a linear relation between basin depth and  
``basin shape''' ${\cal V}$, i.e.,
\begin{equation}
{\cal V}=a(V)+b(V) e_{IS},
\label{eq:shape}
\end{equation}
correctly describes the simulation data (Fig.~\ref{fig3}-(a)).
\section{The statistical properties of the landscape}
\label{sec:sconf}
In the previous section we have discussed how to relate the
basin shape to the basin depth. In this section, we exploit the
formulation of the liquid free energy in the IS formalism to
evaluate the number of PEL basins as a function of the basin depth. 

This quantity is of primary interest both for
comparing with the recent theoretical calculations~\cite{mezpar,coluzzi} 
and to examine some of the proposed relation between dynamics and
thermodynamics~\cite{adam65,schultz,wolynes}
 connecting a purely dynamical
quantity like the diffusion coefficient to a purely thermodynamical
quantity ($S_{conf}$). The number of basins $\Omega$ as a function of 
the basin depth  $e_{IS}$ has  been recently evaluated for a few 
models~\cite{speedymolphys,skt99,scala00,starr01,sastry01,voivod01,speedyjpc,otplungo}
and from the analysis of
experimental data~\cite{stillinger98,richert98,speedyjpc2}.
$S_{conf}(T,V)$ --- the logarithm of the
number of explored basins ---  can be calculated as difference of 
the entropic part of Eq.~(\ref{eq:freeenergy2}) and  
Eq.~(\ref{eq:svib}), i.e., as
\begin{equation}
S_{conf} (T,V)=S(T,V) - S_{vib} (T,V).
\label{eq:sconf}
\end{equation}
In the thermodynamic limit, when fluctuations are negligible, 
a parametric plot in $T$ of $S_{conf}(T,V)/k_B$ 
vs $\langle e_{IS}(T,V)\rangle$ provides an accurate estimate of the
number of basins of depth $e_{IS}$. 
This information, together with the information of the
$e_{IS}$ dependence of the basin shape (or volume) 
(Eq.~(\ref{eq:shape})) completely
defines the statistical properties of the landscape, at least in the
range of $e_{IS}$ values sampled by the system in the studied $T$
region. The availability of $S_{conf}(e_{IS})$ (Fig.~\ref{fig3}-(b))
and ${\cal V}(e_{IS})$  (Fig. ~\ref{fig3}-(a))
opens the  possibility of a modellization of the
thermodynamic of the system in terms of landscape properties, 
as discussed in the next section.
\section{The Random Energy Model: the Gaussian landscape}
\label{sec:rem}
A  modellization of the statistical properties of the landscape
is the next conceptual step in the development of a 
thermodynamic description of the liquid in the IS formalism.
A possible modellization, which appears to be consistent with the
numerical evidence for fragile liquids, is based on the
hypothesis that the the number $\Omega(e_{IS}) de_{IS}$ 
of distinct basins of depth 
between $e_{IS}$  and  $e_{IS}+de_{IS}$ 
in a system of $N$ atoms or molecules is 
described by a Gaussian distribution~\cite{REM,heuer00,starr01,sastry01,press}, 
i.e.,
\begin{equation}
\Omega(e_{IS})de_{IS}=e^{\alpha N}
\frac{e^{-(e_{IS}-E_o)^2/2\sigma^2}}{(2 \pi \sigma^2)^{1/2}}de_{IS}.
\label{eq:gdos}
\end{equation}
Here the amplitude $e^{\alpha N}$ accounts for the total number of
basins, $E_o$ has the role of energy scale and $\sigma^2$
measure the width of the distribution.
One can understand the origin of such distribution 
invoking the central limit theorem. Indeed, 
in the absence of a diverging correlation length, 
in the thermodynamic limit,  each $IS$
can  be decomposed in a sum of independent subsystems, 
each of them characterized by its own value of $e_{IS}$. The system
IS energy, in this case, will be distributed according to 
Eq.~(\ref{eq:gdos}).

Within the two assumptions of Eq.~(\ref{eq:gdos}) ---
Gaussian distribution of basin depths --- and  Eq.~(\ref{eq:shape}) 
---linear dependence of the basin free energy on  $e_{IS}$ ---
an exact evaluation of the partition function can be carried out.
The corresponding Helmholtz free energy is given by~\cite{sastry01}
\begin{eqnarray}
F (T,V)&=&-T S_{conf}(T,V)+ \langle e_{IS}(T,V)\rangle 
+ f_{vib}(E',T) \label{eq:freeenergy3}\\
&+& k_B T b(V)(<e_{IS}(T,V)>-E').\nonumber
\end{eqnarray}
Here $E'$ is an arbitrarily fixed energy,  
\begin{equation}
\langle e_{IS}(T)\rangle =(E_o(V)-b(V)\sigma^2(V))-\beta \sigma^2(V),
\label{eq:eis}
\end{equation}
and
\begin{equation}
S_{conf}(T)/k_B=\alpha(V) N
-(\langle e_{IS}(T,V)\rangle -E_o(V))^2/2
\sigma^2(V).
\label{eq:sconf2}
\end{equation}
Note that, from a plot of $\langle e_{IS}(T)\rangle$ vs $1/T$, 
one can immediately evaluate two of the parameters of the gaussian
distribution, $\sigma^2$ (from the slope) and
$E_o$ (from the intercept). Similarly, from fitting 
$S_{conf}(T)$ according to Eq.~(\ref{eq:sconf2}), one can evaluate
the last parameter $\alpha$ (see Fig.~\ref{fig3}).  

The fitting parameters $\alpha(V)$, $E_o(V)$, and $\sigma^2(V)$ 
depend in general on the volume. A study of the volume 
dependence of these parameters, associated to the $V$-dependence 
of the shape indicators ($a$ and $b$ in Eq.~(\ref{eq:shape}))
provides a full characterization of the volume dependence of the
landscape properties of a model, and offers the possibility of
developing a full equation of state based on 
statistical properties of the landscape.

When comparing numerical simulation data and theoretical predictions,
---Eqs.~(\ref{eq:eis}) and~(\ref{eq:sconf2})--- the range of temperatures 
must be chosen with great care. Indeed, at high $T$,
the harmonic approximation will overestimate the
volume in configuration space associated to an inherent structure.
While in harmonic approximation such a  
quantity is unbounded, the real basin volume is not.
Indeed, the sum of all basins volumes is equal to the
volume of the system in configuration space. Anharmonic corrections, 
if properly handled, should compensate such overestimate, but 
at the present time, no model has been developed to
correctly describe the high $T$ limit of the anharmonic component.  
Numerical studies have shown that
the range of validity of the present estimates of the
anharmonic correction does not extend beyond the temperatures
at which the system shows already a clear two-step relaxation behavior
in the dynamics. Indeed, the presence of a  two-step relaxation 
is a signature that the
system spends a time larger than the microscopic characteristic times
around a well defined local minima.
\section{Conclusions}
In this paper we have discussed  the numerical techniques
employed to  evaluate the statistical properties of the PEL
for molecular systems.  These numerical calculations 
are limited to the region of temperatures and volumes
where equilibrium configurations can be numerically generated.
Still, very simple arguments can be presented which allow to
generalize the results and formulate a full thermodynamic description
of the supercooled liquid state, only  in terms of the statistical 
properties of the PEL.

The possibility of partitioning the free energy and its
thermodynamic derivatives as a sum of configurational and
vibrational degrees of freedom has been recently exploited
to derive a satisfactory  description of the equation of
state~\cite{deben01,press} for supercooled liquids only 
in terms of PEL properties. 
A better understanding of the nature
of each contribution (configurational and vibrational) 
to quantities as the total pressure of the system is achieved.
At the same time, the availability of detailed estimates for the
landscape properties strongly suggests a generalization of this approach
to out of equilibrium conditions. It 
has been  recently shown~\cite{parrocchia} that if the system 
ages exploring the same basins  visited in equilibrium, 
it is possible to write an out of equilibrium equation of state 
expressing $P$ as a function not only of $V$ and $T$ but also 
as a function of the (time dependent) depth of the explored basin.

The availability of numerical estimates for the statistical 
properties of the PEL in models of simple liquids 
should encourage theoreticians to develop schemes for the analytic
evaluation of these quantities. If this goal
were reached,  the understanding the thermodynamics of supercooled liquids
and glasses would improve significantly. 
\section*{Acknowledgments}
We acknowledge support from MIUR-COFIN 2000 and  FIRB.

\vskip 1cm
%

%
%
\end{document}